\documentclass{article}

\usepackage{arxiv}
\usepackage{graphicx}
\usepackage{dcolumn}
\usepackage{bm}
\usepackage[utf8]{inputenc}

\usepackage{natbib}
\usepackage{mathtools}
\usepackage{amsmath}
\usepackage{amssymb}
\usepackage{bm,bbm}
\usepackage{float}
\usepackage{color}
\usepackage{xcolor}
\usepackage{enumerate}

\usepackage{enumitem}

\usepackage{tikz}
\usepackage{comment}

\usepackage{url} 

\newcommand{\beq}{\begin{equation}}
\newcommand{\eeq}{\end{equation}}

\newcommand{\Rel}{{\rm Re}_\lambda}
\newcommand{\eps}{\varepsilon}
\renewcommand{\vec}[1]{\bm{#1}}
\title{TURB-Hel: an open-access  database of helically forced homogeneous and isotropic  turbulence}

\author{
  Luca Biferale \\
  Dept. Physics and INFN\\
  University of Rome Tor Vergata and INFN, Italy\\
  \texttt{biferale@roma2.infn.it} \\
\And
  F. Bonaccorso \\
  Dept. Physics and INFN\\
  University of Rome Tor Vergata, Italy.\\
  \texttt{fabio.bonaccorso@roma2.infn.it} \\
\And
  Moritz Linkmann \\
  School of Mathematics and Maxwell Institute for \\
     Mathematical Sciences \\
  University of Edinburgh, UK \\
  \texttt{moritz.linkmann@ed.ac.uk} \\
\And
  Damiano Capocci \\
  Dept. Physics and INFN \\
  University of Rome Tor Vergata, Italy.\\
  \texttt{damiano.capocci@roma2.infn.it}
}

\begin{document}
\maketitle

\begin{abstract}
We present TURB-Hel, a database formed by two datasets of incompressible homogeneous and isotropic turbulence, maintained in a statistically stationary state by fully helical forcing. The aim is to provide a dataset that clearly exhibits the phenomenon of the helicity cascade from the large to the small scales generated by a large-scale forcing that breaks the mirror symmetry. This database offers the possibility to realize a wide variety of analyses of fully developed turbulence from the sub-grid scale filtering up to the validation of an \emph{a posteriori} LES.\\
TURB-Hel is available for download using the SMART-Turb portal http://smart-turb.roma2.infn.it.
\end{abstract}

\section{Introduction}
%We present two datasets of incompressible homogeneous and isotropic 
%turbulence, maintained in a statistically stationary state by fully helical forcing. 
Kinetic helicity, the correlation between velocity fluctuations, $\bm{u}$, and vorticity fluctuations, $\boldsymbol{\omega}$, is known to influence turbulent dynamics and coherent structures formation \cite{waleffe1992} (other references) and to be connected to the breaking of the symmetry under inversion of all axes in a rotationally invariant ensemble.\\
High levels of helicity have been associated with: a strong depletion of the kinetic energy fluxes across scales \cite{kraichnan1973} as a consequence of the nonlinearity reduction caused by the alignment of velocity and vorticity and with low dissipation regions \cite{moffatt2014}.
Like energy, helicity is a quadratic invariant of the Euler equations but, unlike energy, it is not sign definite. In this respect, the interactions between helical Fourier modes of opposite signs (heterochiral sector) contribute to the energy transfer from large to small scales \cite{waleffe1992} \cite{alexakis2018} \cite{alexakis2017} while those with the same sign (homochiral sector) are responsible for the inverse energy transfer across scales in the inertial range \cite{biferale2012}, \cite{biferale2013}, \cite{Sahoo15}.

Any solenoidal vector field can be decomposed into positively and negatively helical components \cite{waleffe1992}. That is, the velocity field can we written as
\begin{equation}
    \bm{u}(\bm{x},t) =\bm{u}^+(\bm{x},t) + \bm{u}^-(\bm{x},t) \ .
\end{equation}
The fields $\bm{u}^+$ and $\bm{u}^-$ are obtained 
by projecting the Fourier coefficients $\hat{\bm{u}}(\bm{k},t)$ onto basis vectors which are eigenfunctions of the curl operator in Fourier space
\begin{equation}
    \hat{\bm{u}}^\pm(\bm{k},t) = u^\pm (\bm{k},t) \bm{h}^\pm(\bm{k}) \ , 
\end{equation}
where $i \vec{k} \times k\bm{h}^\pm(\bm{k}) = \pm \bm{h}^\pm(\bm{k})$. The total kinetic helicity can then be calculated as 
%\begin{align}
%    \langle \bm{u}(\bm{x},t) \cdot \boldsymbol{\omega}(\bm{x},t)\rangle_V 
%    & = \sum_{\bm{k}} \hat{\bm{u}}(\bm{k},t) \cdot \hat{\boldsymbol{\omega}}(\bm{k},t) \nonumber \\
%    &= \sum_{\bm{k}} \left (u^+(\bm{k},t)\bm{h}^+(\bm{k}) + u^-(\bm{k},t)\bm{h}^-(\bm{k})\right)
%    \cdot k\left (u^+(\bm{k},t)\bm{h}^+(\bm{k}) - u^-(\bm{k},t)\bm{h}^-(\bm{k})  \right) \nonumber \\
%    & = \sum_{\bm{k}} k \left( u^+(\bm{k},t)^2 - u^-(\bm{k},t)^2 \right) \ ,
%\end{align}
\begin{equation}
    \langle \bm{u}(\bm{x},t) \cdot \boldsymbol{\omega}(\bm{x},t)\rangle_V 
     = \sum_{\bm{k}} \hat{\bm{u}}(\bm{k},t) \cdot \hat{\boldsymbol{\omega}}(\bm{k},t) 
   = \sum_{\bm{k}} k \left( u^+(\bm{k},t)^2 - u^-(\bm{k},t)^2 \right) \ ,
\end{equation}
where $\langle \cdot \rangle_V$ denotes a spatial average.

\section{Numerical simulations}\label{sec:numerics}
Data has been generated by direct numerical simulation of the 3D Navier-Stokes equations
on a triply periodic domain of size $L_{\rm box} = 2\pi$ in each spatial direction
\begin{align}
	\label{eq:momentum}
	\partial_t \bm{u} +\bm{u} \cdot \nabla \bm{u} & = - \nabla p + \nu \Delta \bm{u} + \bm{f} \ , \\
	\label{eq:incomp}
	\nabla \cdot \bm{u} & = 0 \ ,
\end{align}
where $\bm{u}$ denotes the velocity field, $p$ is the pressure divided by 
the fluid density, which is constant, and $\nu$ the kinematic viscosity. The forcing $\bm{f}$ is a random Gaussian process with zero mean 
active in the wavenumber band $k \in [0.5,2.4]$ for the dataset HL2 while for HL1 the forcing ranges in the interval $k \in [1.0,2.0]$. It satisfies $\nabla \cdot \bm{f} = 0$, and is fully helical, $\bm{f} = \bm{f}^+$. 
Equations \eqref{eq:momentum}-\eqref{eq:incomp} were stepped forwards in time using a second-order Adams-Bashforth scheme, with the viscous term treated implicitly using an integrating factor. The spatial discretisation was implemented via the standard pseudospectral method with complete dealiasing by the two-thirds rule. Further details and mean values of key observables are summarised in table \ref{tab:simulations}.

\begin{table}[H]
  \begin{center}
\def~{\hphantom{0}}
   \begin{tabular}{ccccccccccccc}
        \hline
        \hline
		 id &$N$ & $E$  & $\nu$ & $\eps$ & $L$ & $\tau$  & $\Rel$ &  $\eta/10^{-3}$ & $k_{\rm max}$ & $k_{\rm max} \eta$ & $\Delta t / \tau$ & \# \\
   \hline
		 HL1 & 512 & 6.32  & 0.002 & 2.88   & 1.14  & 0.50   & 240    &  7.41  & 169  & 1.25 & 0.60  & 56 \\
        \hline
		HL2 & 1024 & 7.26  & 0.001 & 3.33   & 1.12  & 0.50    & 327    & 4.20  & 340  & 1.43 & 0.60  & 39 \\

        \hline
      \hline
        \end{tabular}
        \caption{Simulation parameters and key observables, where 
        $N$ denotes the number of collocation points in each coordinate,
        $E$ the total kinetic energy per unit volume,
        $\nu$ the kinematic viscosity,
        $\eps$ the dissipation rate,
        $L = (3 \pi/4 E) \sum_{k = 1}^{k_{\rm max}}  \ E(k)/k \ \Delta k$ the integral scale ,
        $\tau = L/\sqrt{2E/3}$ the large-eddy turnover time,
        $\Rel$ the Taylor-scale Reynolds number,
        $\eta = (\nu^3/\eps)^{1/4}$
        the Kolmogorov microscale,
        $k_{\rm max}$ the largest wave number after de-aliasing,
        $\Delta t$ the ampling interval which is calculated from
        the length of the averaging interval divided by the number of equispaced snapshots, and 
	$\#$ the number of snapshots.   
	The higher-resolved data corresponds to run 22 of \cite{Sahoo15}. 
	Both datasets have been discussed in \cite{us2023} and are  available for download using the SMART-Turb portal { \tt http://smart-turb.roma2.infn.it}. 
	}
  \label{tab:simulations}
  \end{center}
\end{table}

\begin{figure}[htbp]
         \includegraphics[width=.49\columnwidth]{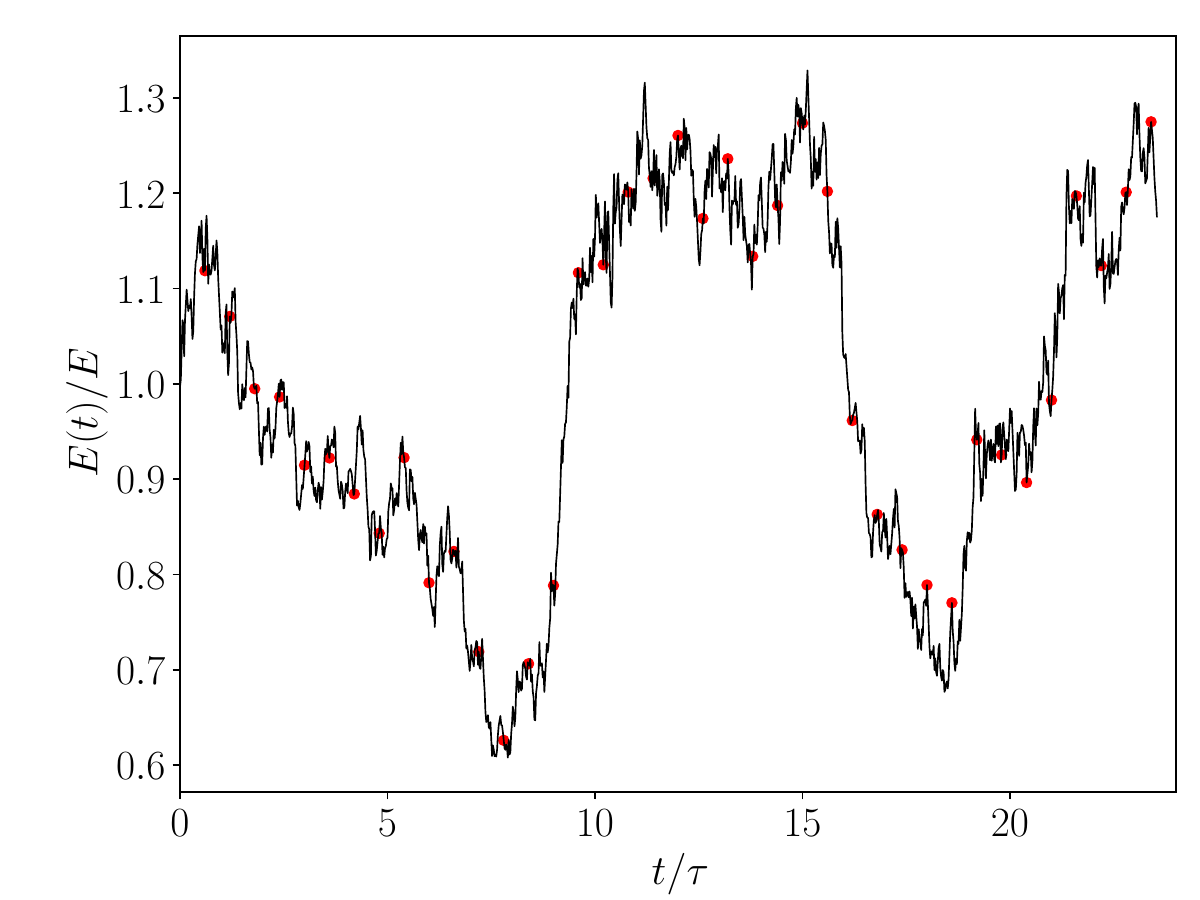}
         \includegraphics[width=.49\columnwidth]{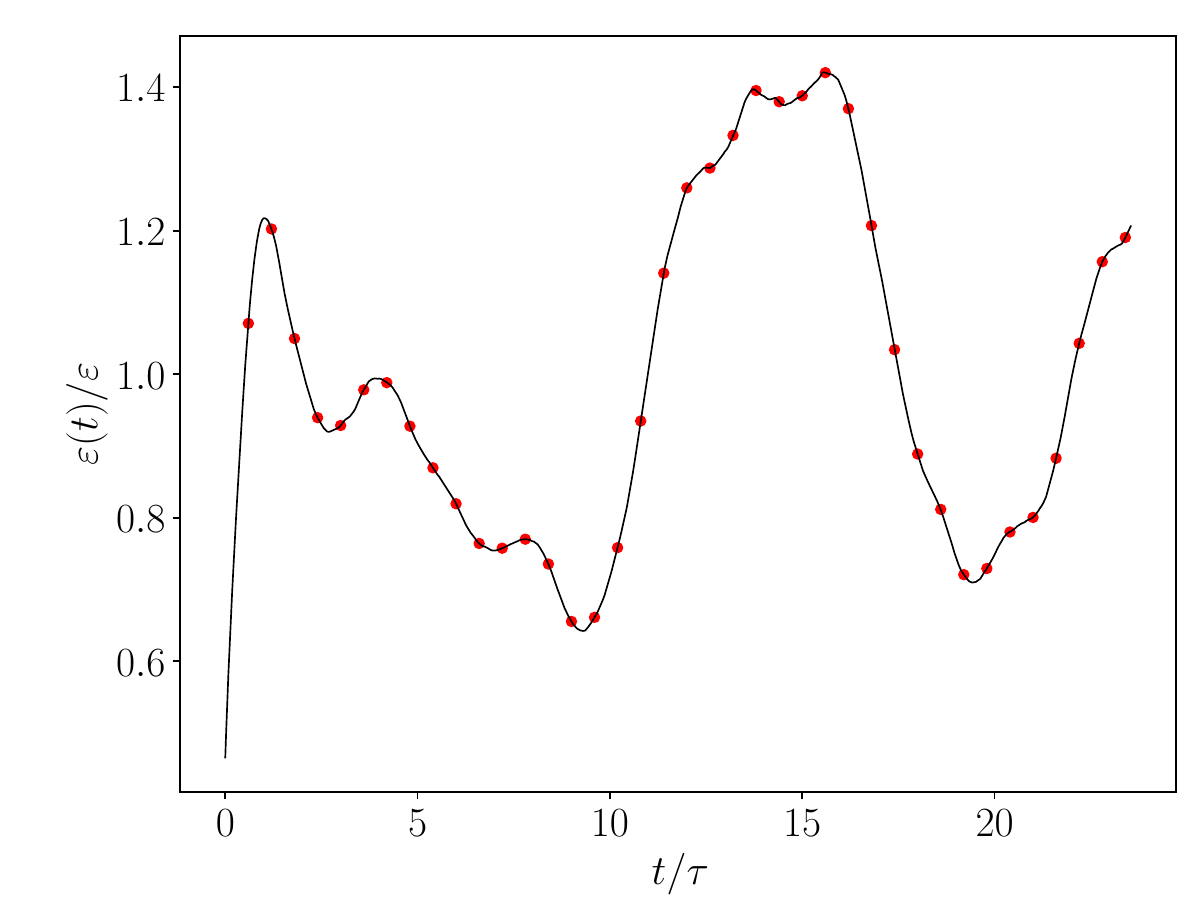}
         \includegraphics[width=.49\columnwidth]{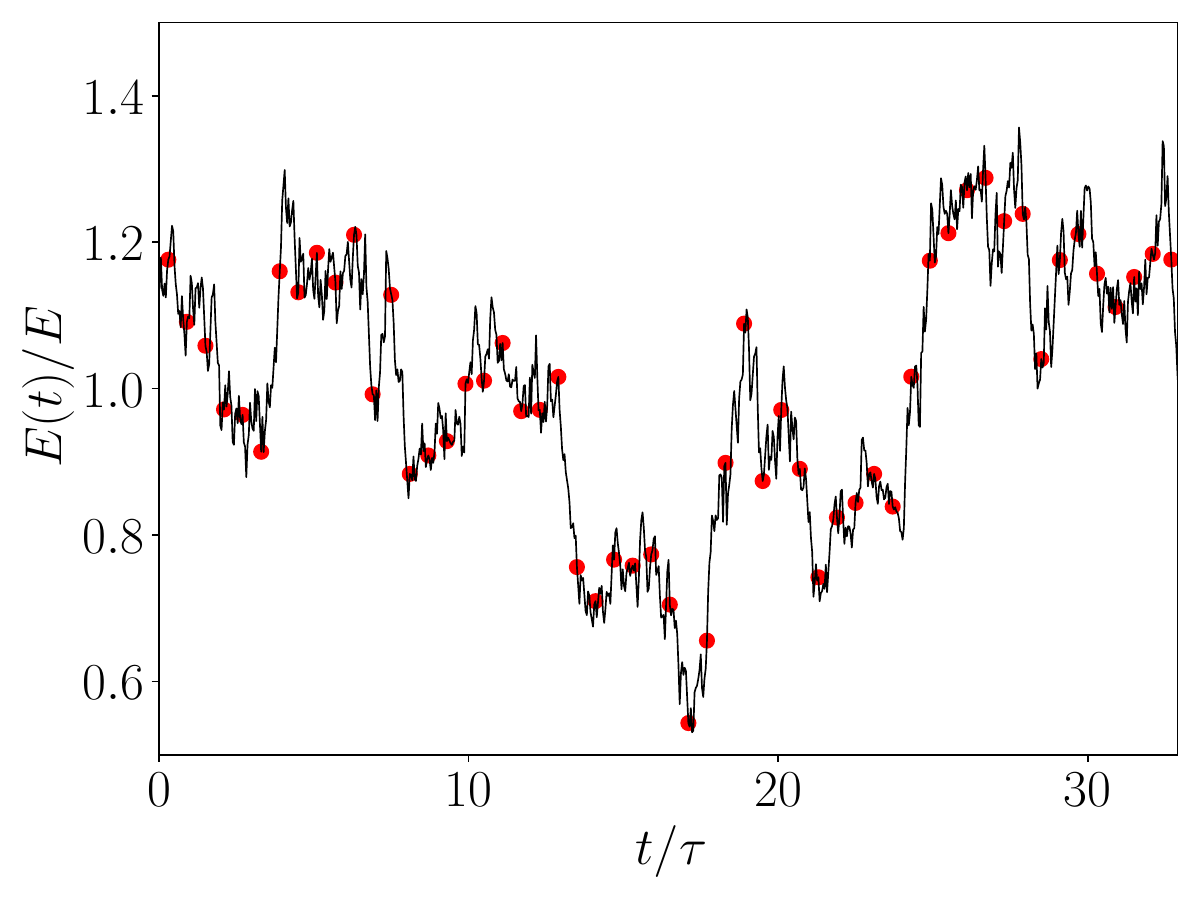}
         \hspace{0.38cm}
         \includegraphics[width=.49\columnwidth]{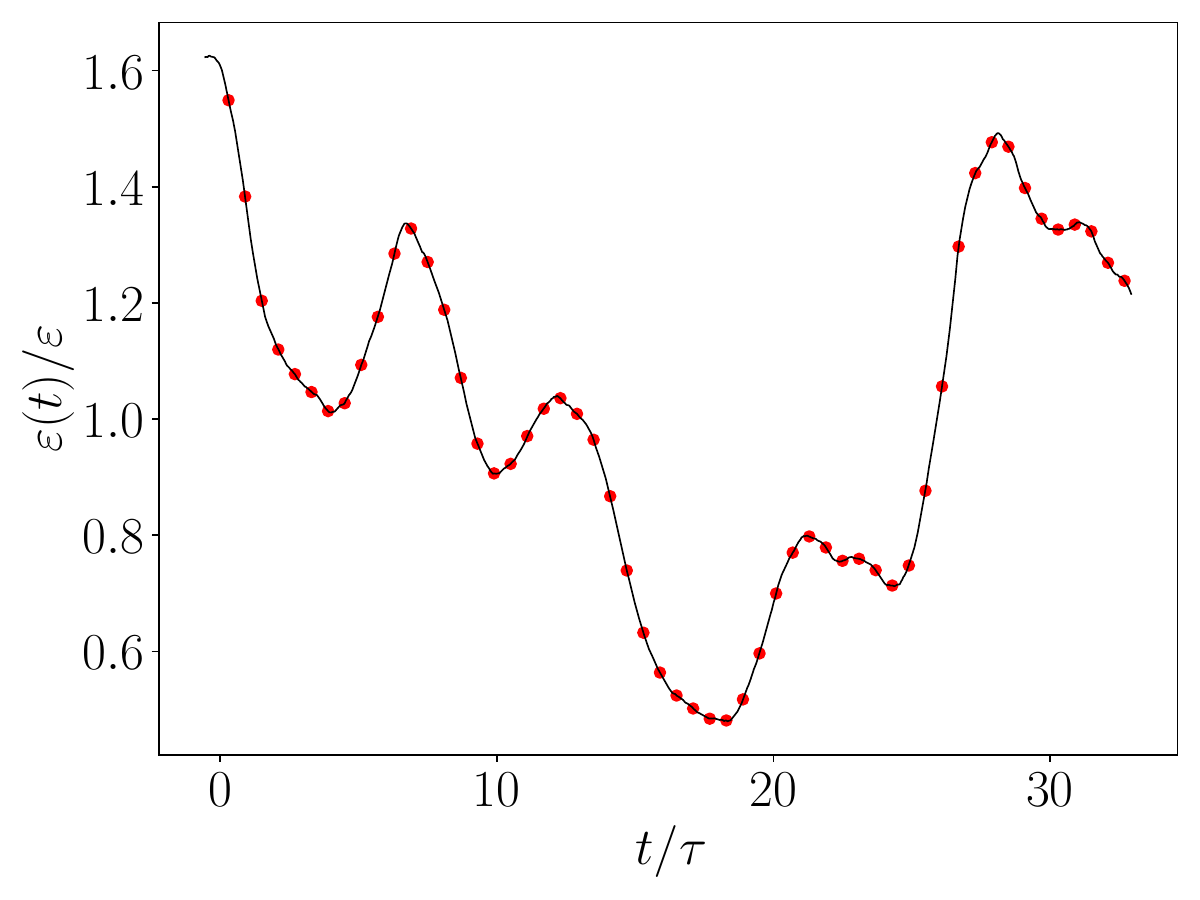}
	 \caption{Time evolution of the total energy (left) and the total dissipation (right).
         The red dots correspond to the sampled velocity-field configurations.\\
         Top panel: dataset HL2. \\
         Bottom panel: dataset HL1.
         }
	 \label{fig:timeseries}
\end{figure}

Time series of the total kinetic energy per unit volume, $E(t) = \frac{1}{2}\langle |\bm{u}(\bm{x},t)|^2 \rangle_V$, and the total dissipation, $\varepsilon(t) = \langle |\nabla \bm{u}(\bm{x},t)|^2 \rangle_V$, for datasets HL1 and HL2, are presented in Fig.~\ref{fig:timeseries}. The red points correspond to instances in time where full data cubes have been sampled.

\begin{figure}[htbp]
        \hspace{-1.2cm}
        \includegraphics[width=.6\columnwidth]{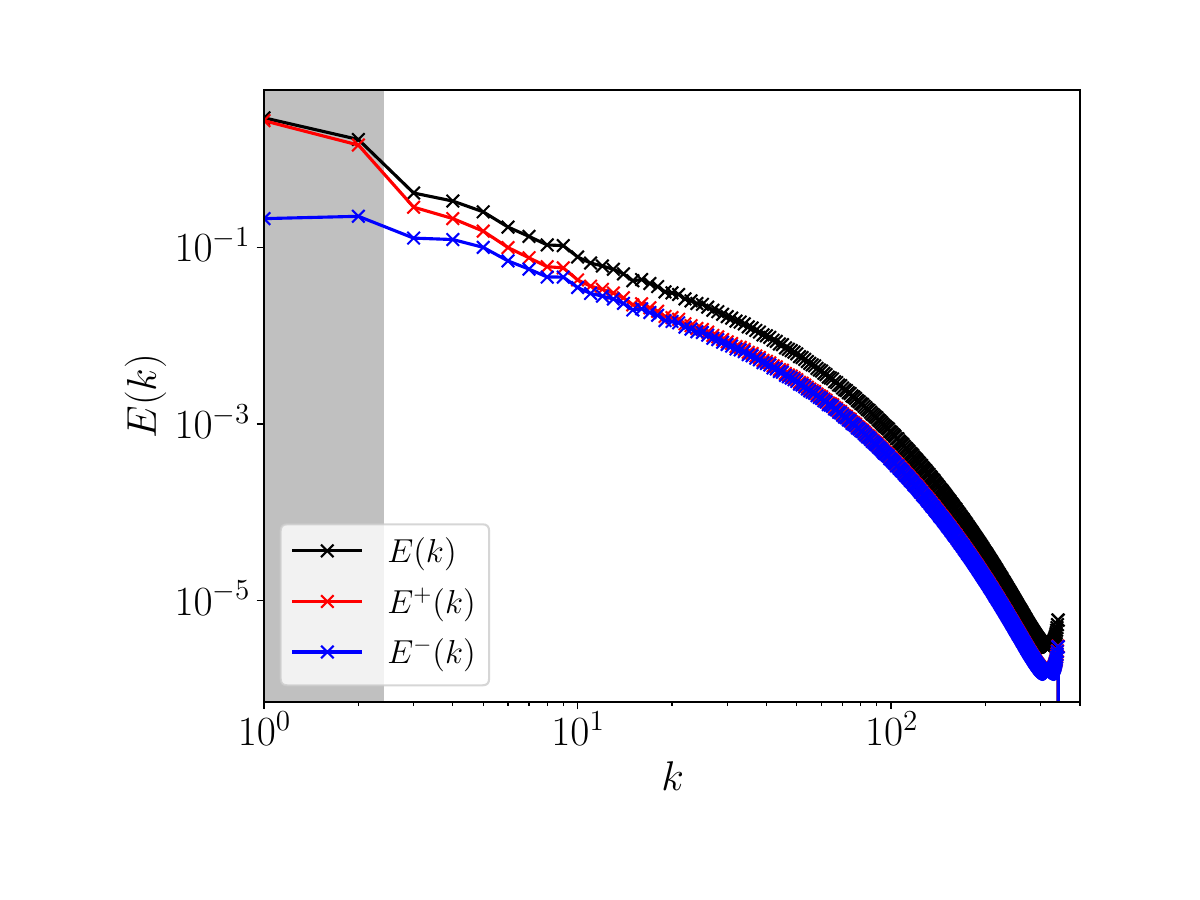}
        \includegraphics[width=.6\columnwidth]{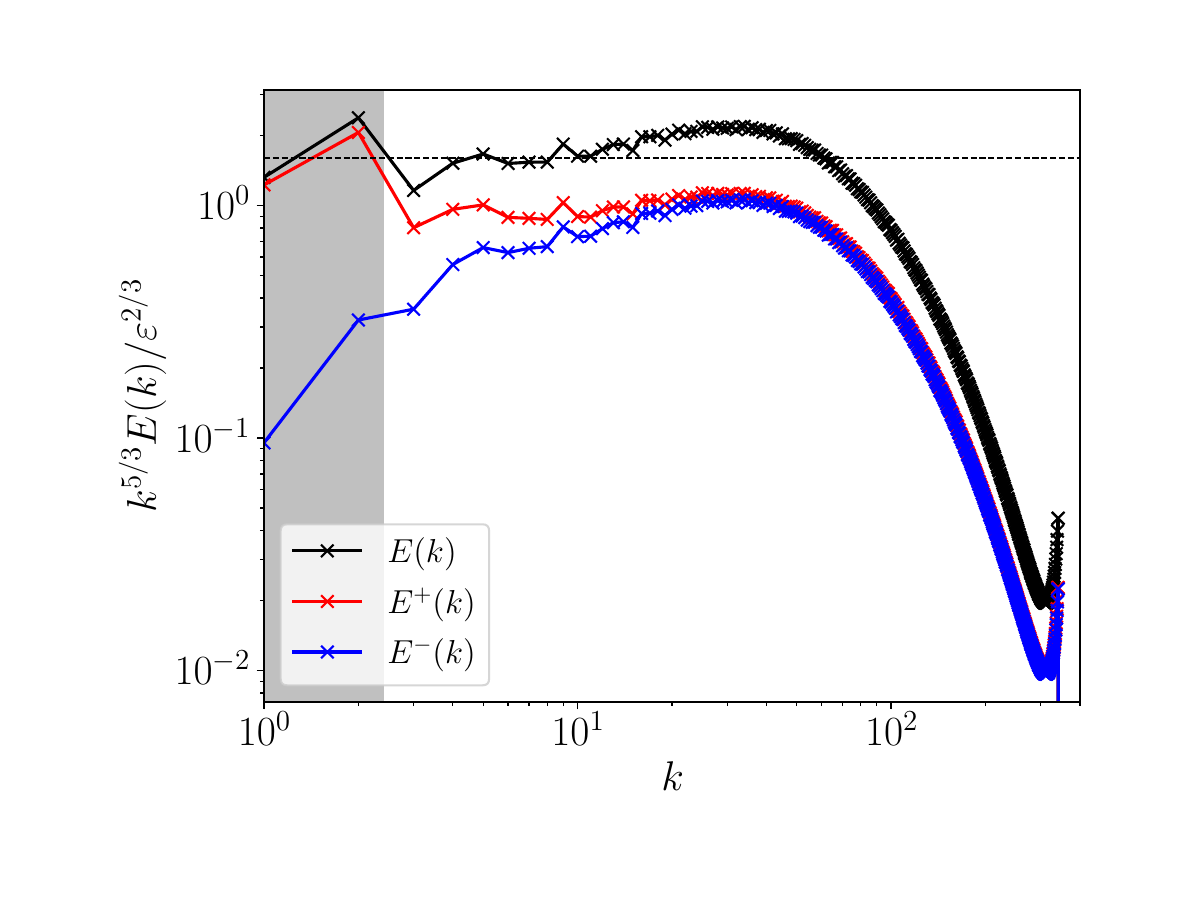}
        \hspace{4.5cm}
        \includegraphics[width=.48\columnwidth]{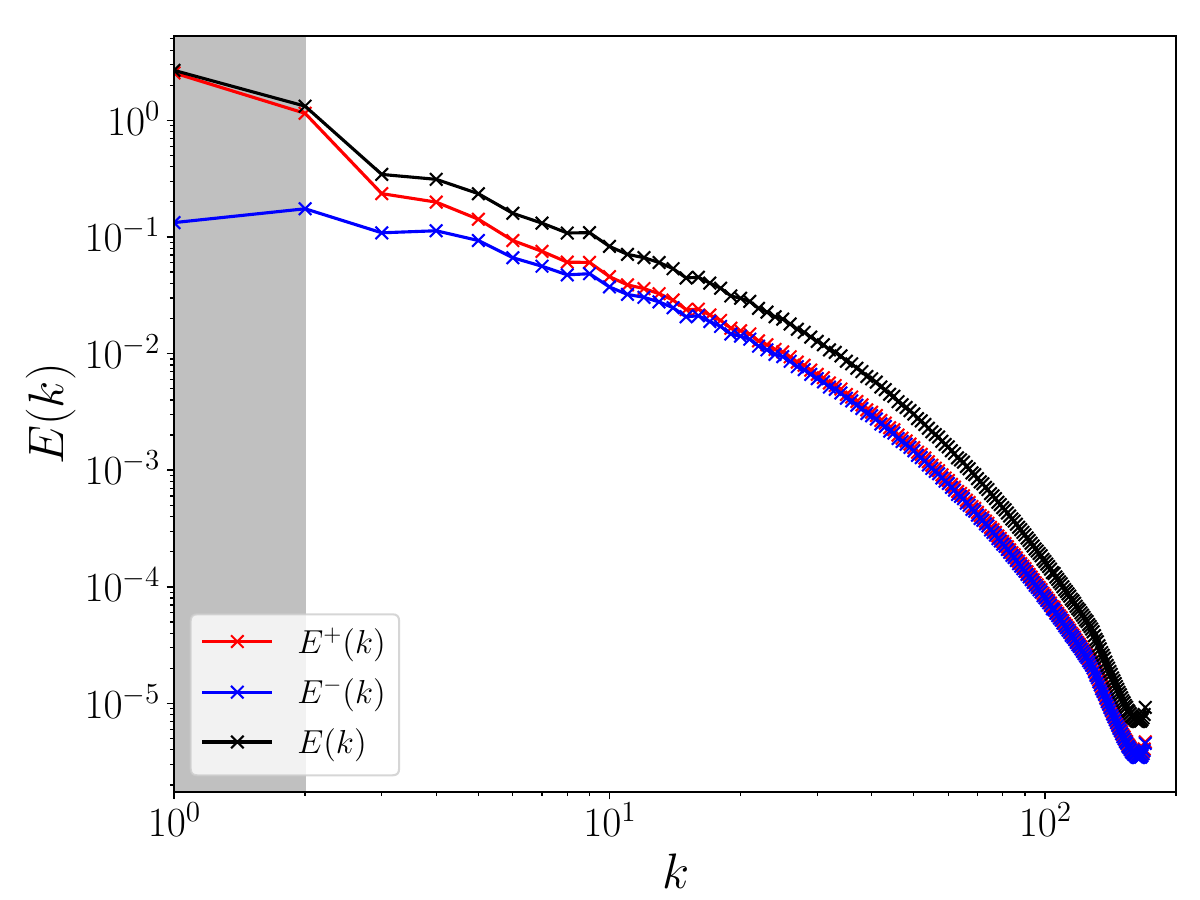}
        \hspace{1.8cm}
        \includegraphics[width=.48\columnwidth]{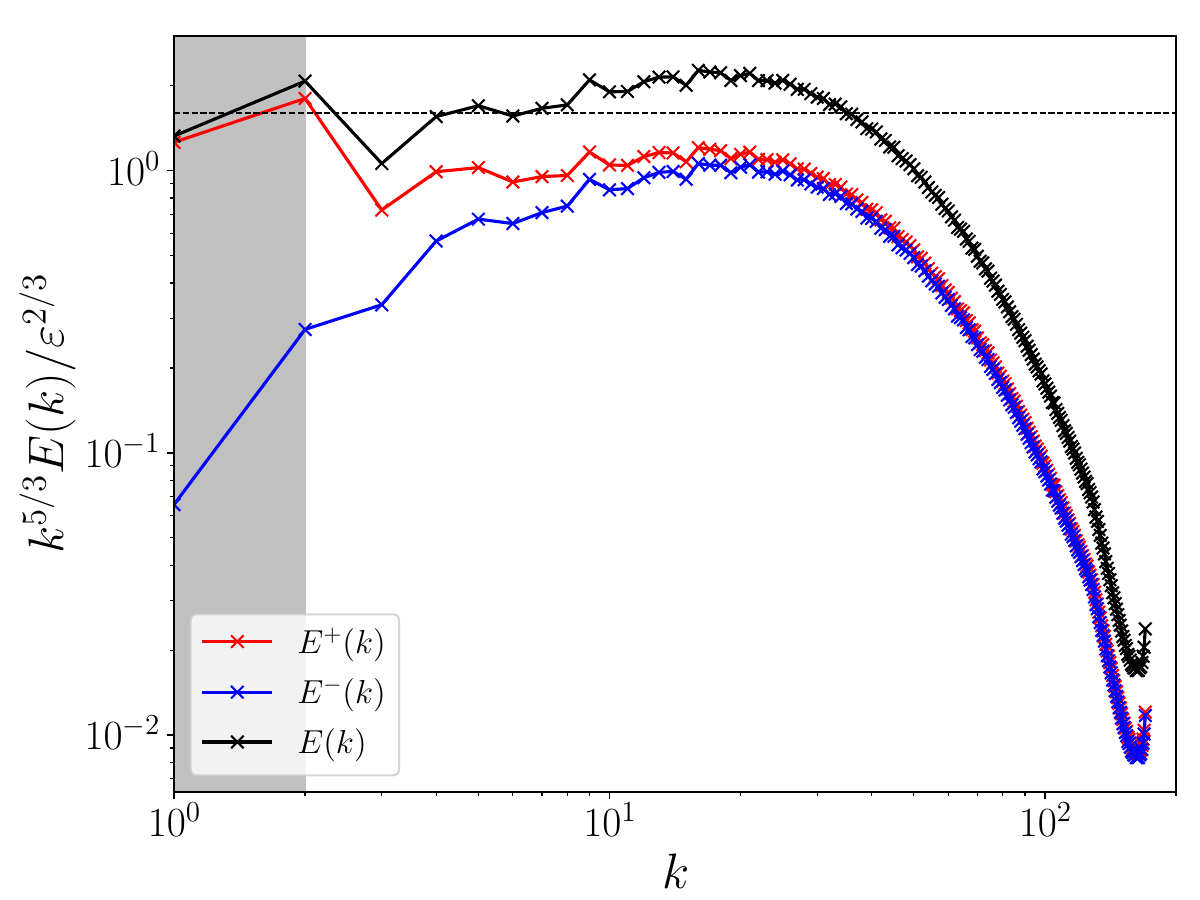}
	\caption{Time-averaged energy spectra, in raw (left) and Kolmogorov-compensated (right) form. 
        The grey-shaded area indicates the forcing range, the positively helical component is shown in red, the negatively helical component in blue 
	and the total energy spectrum in black.  \\
Top panel: dataset HL2. \\
Bottom panel: dataset HL1: a fewer amount of wavenumbers is the consequence of the reduction in the grid point.
In both datasets the dashed lines in the right panels indicate a Kolmogorov constant $C_K \approx 1.6$.}
	\label{fig:spectra}
\end{figure}

Figure \ref{fig:spectra} shows the energy spectrum $E(k) = \frac{1}{2}\sum_{k \leqslant |\bm{k}| < k + 1} |\hat{\bm{u}}(\bm{k})|^2$ averaged over the ensemble of 56 and 39 velocity-field configurations that comprise the datasets of HL1 and HL2 respectively, in raw and Kolmogorov-compensated form. The energy spectrum can be decomposed into contributions from the positively and negatively helical components of the velocity field, $E(k) = E^+(k) + E^-(k)$. In both panels of the figure, the black lines correspond to $E(k)$, the red lines to the energy spectrum of the positively helical velocity-field fluctuations, $E^+(k)$ and the blue lines to that of the negatively helical velocity-field fluctuations, $E^-(k)$. The grey-shaded area identifies the forcing range, and the horizontal dashed line, in the compensated spectra of both panels, indicates a Kolmogorov constant $C_K = 1.6$. 
As the forcing is positively helical, the negatively helical fluctuations must be maintained by nonlinear coupling to the positively helical fluctuations. As can be seen from the date presented in Fig.~\ref{fig:spectra}, the velocity field is positively helical at and near the forcing scales.  
In both datasets, the mirror symmetry is restored towards the end of the inertial range, around $k = 80$. Only the $E(k)$ and $E^+(k)$ scale approximately as $k^{-5/3}$, while $E^-(k)$ has a shallower slope.\\

\section{DataBase Description}
TURB-Hel database is made of files extracted from the $512^3$ HL1 and $1024^3$ HL2 simulation described in the previous section as follows:

\begin{itemize}
    \item During the simulation we dumped the vector potential of the velocity field in fourier space every $3000 dt$ simulation steps. In total we stored 56 (HL1) and 39 (HL2) configurations, we call them "cb" (from Complex B-field);

    \item In order to recover the velocity fields, every configuration should be read using the HDF5 library, then a curl operation is needed to compute the velocity field in Fourier space.

    \item If the velocity field is needed in real space, a backward FFT is needed.

    \item In the support materials, there is a C program \begin{verbatim} read_cb \end{verbatim} which performs all the above steps, and an accompanying ReadME.pdf for the details on how to compile it. 
\end{itemize}

The database TURB-Hel is available for download using the SMART-Turb portal at \url{http://smart-turb.roma2.infn.it}. 
\\

Other data-sets concerning rotating turbulence (TURB-Rot \cite{TURB-Rot}) and Lagrangian particle under turbulence (TURB-Lagr) are already available.  \\

\section*{Acknowledgements}
This work received funding from the European Research Council (ERC) under the
European Union's Horizon 2020 research and innovation programme (grant
agreement No 882340).

\clearpage

\bibliographystyle{jfm}
\bibliography{references}

\end{document}